\magnification=1200
\hsize=17 true cm
\vsize=22 true cm
\tolerance=6000
\baselineskip=15pt
\def\ref {\par \noindent \parshape=6 0cm 14.5cm
 0.5cm 14.5cm 0.5cm 14.5cm 0.5cm 14.5cm 0.5cm 14.5cm 0.5cm 14.5cm}

\newbox\grsign \setbox\grsign=\hbox{$>$} \newdimen\grdimen \grdimen=\ht\grsign
\newbox\simlessbox \newbox\simgreatbox
\setbox\simgreatbox=\hbox{\raise.5ex\hbox{$>$}\llap
     {\lower.5ex\hbox{$\sim$}}}\ht1=\grdimen\dp1=0pt
\setbox\simlessbox=\hbox{\raise.5ex\hbox{$<$}\llap
     {\lower.5ex\hbox{$\sim$}}}\ht2=\grdimen\dp2=0pt

\def\simless{\mathrel{\copy\simlessbox}}
\newbox\simppropto
\setbox\simppropto=\hbox{\raise.5ex\hbox{$\sim$}\llap
     {\lower.5ex\hbox{$\propto$}}}\ht2=\grdimen\dp2=0pt

\vtop to 2.5 true cm {}
\centerline {\bf THE NEAR-INFRARED NAI DOUBLET FEATURE IN M STARS
\footnote*{\rm Observations collected at the European
 Southern Observatory, ESO, La Silla, Chile}}

\vskip 3.5 true cm
\centerline {R.P. Schiavon, B. Barbuy, S.C.F. Rossi, A. Milone}
\bigskip
\centerline { Universidade de S\~ao Paulo, IAG, Dept. de Astronomia,}
\centerline { CP 9638, S\~ao Paulo 01065-970, Brazil}
\vskip 5. true cm
\noindent Submitted to: The Astrophysical Journal
\bigskip
\noindent Send proofs to: R. P. Schiavon

\vfill\eject

\noindent {\bf Abstract.} 
 The NaI near-infrared feature has been used to indicate the dwarf/giant
population in composite systems, but its interpretation
still is a contentious issue.
In order to try to understand the behavior of this controversial feature, we
study the spectra of cool stars, by means of both observed and
synthetic spectra.

We conclude that the NaI infrared feature can be used
as a dwarf/giant indicator. We propose a modified definition
of the NaI index by defining a red continuum at $\lambda$8234 ${\rm \AA}$ 
and by measuring 
 the equivalent width in the range 
8172 - 8197 ${\rm \AA}$, avoiding the region
at $\lambda$ $>$ 8197 ${\rm \AA}$ which contains
 VI, ZrI, FeI and TiO lines. 

\bigskip

\bigskip
\noindent {\bf Subject Headings:} stars: atmospheres, 
galaxies: stellar content, spectroscopy
\bigskip

\bigskip
\noindent {\bf 1. Introduction}
\bigskip
\noindent The contribution of M dwarfs to the integrated spectra
of galaxies  still is a matter of debate. In the near-infrared, one
may expect to detect most of the contribution from M-type stars to the
integrated light of galaxies. Investigations in this spectral range
have involved a number of gravity sensitive indices, such as the NaI
doublet ($\lambda\lambda$ 8183 and 8195${\rm \AA)}$, the Wing-Ford band of FeH
($\lambda$ 9900${\rm \AA}$), the CaII triplet ($\lambda\lambda$ 8498, 8542 and 
8662${\rm \AA}$) and the CO band at 2.3$\mu m$.
Based on observations of the 
 NaI doublet at $\lambda$8183, $\lambda$8195 ${\rm \AA}$,
Spinrad \& Taylor (1971, hereafter ST)
suggested the presence of a strong
dwarf component in M31 and M81.
Whitford (1977) used the gravity discriminator
FeH Wing-Ford band to conclude that the ST suggestion
could not be confirmed. 
Cohen (1978) observed the
infrared NaI doublet, CaII triplet, FeH and TiO bands in M31 and M32,
concluding that no enhancement of cool dwarf population is required
to explain these features, but that instead,
the effect seen is due to the higher metallicity of M31 relative to M32.
 Faber \& French (1980, hereafter FF80) 
have used the intensity of this feature as an indicator of
dwarf/giant ratio,
returning to the same conclusion of ST: an excess of M dwarfs in M31.
The question was further studied by Persson et al. (1980) through the 
observation of CO and H$_2$O narrow-band indices in M31, having measured
the indices from the bulge to the nucleus; the CO data do not support
a dwarf-enriched nucleus, but the infrared data and the NaI feature are
 consistent with a metallicity increase of a factor 3 between the bulge
and the semistellar nucleus.
\par
 Alloin \& Bica (1989, hereafter AB89) studied the behavior of this feature
in individual stars and in the integrated light of stellar clusters and galaxies.
They concluded that the enhancement of the feature in galaxies is due
to another absorber, possibly a TiO bandhead at $\sim
\lambda$8205 ${\rm \AA}$.
Xu et al. (1989) claimed that this feature
is actually a blend of TiO, FeI and MgI lines 
which are observed in metal-rich systems.
\par Regarding the use of the NaI feature in other external galaxies,
 Boroson \& Thompson (1991) found strong 
evidence that points to a dwarf-dominated light in  NGC 4472,
suggesting that population gradients may be present. 
 Delisle \& Hardy (1992) observed
10 bulges of spirals and ellipticals
in the spectral range 8000-10000 ${\rm \AA}$,
concluding again that the NaI feature is probably a result of a
blend and that a metallicity effect may be present.
Couture \& Hardy (1993) studied the Wing-Ford band
in six galaxies concluding that M dwarfs cannot contribute to
more than 20\% of the I-luminosity of their sample galaxies.
\par Therefore the interpretation of the NaI $\lambda$ 8190 ${\rm \AA}$
feature in galaxies is surrounded by
a controversy, whether it is predominantly a dwarf/giant
ratio indicator or a metallicity sensitive feature.
 In an attempt to solve this question,
 we investigate in detail
the behavior of the NaI feature through the observation
of cool dwarf and giant stars, and
by computing synthetic spectra for a grid of
 stellar parameters,
employing updated
model atmospheres.
\par In Sect. 2 the observations are reported. 
The calculations of synthetic spectra
are described in Sect. 3. In Sect. 4 the
behavior of the NaI feature as a function of stellar
parameters is shown. In Sect. 5 the results are discussed.
\bigskip
\noindent {\bf 2. Observations}
\bigskip
\noindent  The observations were carried out in February/1996 
at the 1.5m and 1.4m telescopes
of the European Southern Observatory (ESO), La Silla, Chile. 
At the 1.5m telescope, the
Boller \& Chivens spectrograph was used with a grating of 1200 l/mm
(ESO grating \# 11), yielding a dispersion of 66 ${\rm \AA}$/mm
and a resolution of $\Delta\lambda$ $\approx$ 2 ${\rm \AA}$
(1 ${\rm \AA}$/pixel). The Ford Aerospace ESO CCD \# 24, of
2048x2048 pixels and pixel size of 15x15 $\mu$m was used, and a
wavelength coverage of $\lambda\lambda$ 8080-10100
${\rm \AA}$ was obtained.
\par The spectral types of the sample stars range from M0 to M7.
The sample stars are reported in Table 1, where the
visual magnitudes, colours, spectral types and temperatures 
are indicated. The temperatures were derived by 
using spectral types vs. temperature calibrations by Fluks et 
al. (1994) for giants, and a (R-I)$_{Cousins}$ vs. temperature calibration by
Bessell (1991) for dwarfs.
\par In Figs. 1a and 1b are displayed  spectra of giants
and dwarfs respectively, ranging in temperature from spectral types
M0 to M7.
\par The giant HR3099 was also observed at high resolution
at the ESO 1.4m Coud\'e Auxiliary Telescope (CAT). 
The Loral/Lesser
CCD ESO \# 38 with 2688x512 pixels, with
pixel size of 15x15 $\mu$m was used, at a resolution of
R = 45000.

\bigskip
\noindent {\bf 3. Calculations}
\bigskip
\noindent The code for spectrum synthesis calculations is 
described in Barbuy (1982), where the computation of molecular
lines was included in the code RAI11 from M. Spite.
The model atmospheres employed are from Kurucz (1992)
for T${\rm eff}$ $>$ 3500 K, Plez et al. (1992)
for M giants of 2500 $<$ T${\rm eff}$ $<$ 3600 K
  and Allard \& Hauschildt (1995) for M dwarfs of 
2500 $<$ T${\rm eff}$ $<$ 3500 K.
 The atomic line list is from Moore et al. (1966) and the
oscillator strengths were obtained through a fit to the solar spectrum.
The NaI doublet is in fact a triplet but two lines are
almost coincident.
The atomic constants adopted for the NaI lines were the following:
log gf = +0.22, -0.479, +0.477, respectively for 
NaI$\lambda$8183.25, $\lambda$8194.79 
and  $\lambda$8194.84 ${\rm \AA}$ 
(Wiese et al. 1969), and interaction constants of 0.3E-30.

\par Molecular lines of 
the TiO A$^3$$\Phi$ - X$^3$$\Delta$ $\gamma$ system
and 
CN A$^2$$\Pi$ - X$^2$$\Sigma$ red system
were included. For TiO the vibrational transitions
in the studied spectral region 
are  (3,4), (4,5) and (0,2),
 where the line lists were 
kindly made available by J.G. Phillips.
For CN, the vibrational transitions 
are (7,4), (8,5), (2,0), (3,1) and (4,1),
 where the lists of rotational lines
were adopted from Davis \& Phillips (1963).
Further details on the  molecular constants used are described in
Erdelyi-Mendes \& Barbuy (1991) and Milone \& Barbuy (1994).
\par In order to verify the reliability of our spectrum synthesis,
we show in Fig. 2 the fit of the synthetic spectrum
computed with (T$_{\rm eff}$, log g, [Fe/H], v$_{\rm t}$ = 
(3100 K, 4.5, 0.0, 1.0 km.s$^{-1}$)
to the observed spectrum of the dwarf GL 493.1.

\bigskip
\noindent {\bf 4. Grid of synthetic spectra}
\bigskip
\noindent We have computed a grid of synthetic spectra in the
wavelength range $\lambda\lambda$ 8080 - 8320 ${\rm \AA}$, for the stellar
parameters reported in Table 2.

\par The equivalent widths of the NaI $\lambda$ 8190 ${\rm \AA}$ feature
were measured in  two wavelength ranges: $\lambda\lambda$ 8172-8209
${\rm \AA}$ (EW$_1$),  as adopted by FF80 and AB89, 
and $\lambda\lambda$ 8172-8197  ${\rm \AA}$ (EW$_2$), where
essentially only the NaI doublet lines are measured.
 The red continua are different in the two measurements:
 the red continuum point adopted by FF80 and AB89 
fall onto the bottom of a TiO band, and for this
reason we used a better defined continuum point at
$\lambda$ 8234 ${\rm \AA}$; the two continua definitions can be seen
 in Fig. 3. 
In Table 3, the wavelengths of NaI indices and continua are given.
Note that the continua defined by FF80 are only given in their
Fig. 2, from which we deduced the ranges given in our Table 3 
for their definitions of EW$_1$.
\par In Figs. 4a and 4b are plotted the equivalent widths of
the NaI feature adopting the NaI index definitions
EW$_1$ and EW$_2$ respectively,
for giants and dwarfs, in the range of temperatures 2500 $<$
T$_{\rm eff}$ $<$ 5500 K.
In these Figs. are also illustrated shifts in equivalent
widths, in the
temperatures of overlap, between the models by 
Allard \& Hauschildt (1995), Plez et al. (1992) and Kurucz (1992).
It is evident from Fig. 4b that the
NaI feature is clearly stronger in dwarfs of T$_{\rm eff}$
 $\simless$ 4000 K, indicating that the feature can in fact
be used as a reliable indicator of the presence of cool dwarfs
in composite systems.

\bigskip
\noindent {\it 4.1 Contaminating features at $\lambda$ 8197-8205 ${\rm \AA}$}
\bigskip
\noindent 
The TiO bandhead of the
vibrational transition (0,2) of the  A$^3$$\Phi$ - X$^3$$\Delta$ 
$\gamma$ system
at $\lambda$8205 ${\rm \AA}$ is clearly present in giants of
T$_{\rm eff}$ $\simless$ 3800 K, becoming increasingly
stronger than the NaI lines for decreasing temperatures.
For such low temperatures we also note that the TiO band
weakens for higher gravities.
\par Atomic lines are also present
 redwards of the $\lambda$8197 ${\rm \AA}$
and they become stronger for cooler stars, as already proposed by
Xu et al. (1989).
\par The high resolution spectrum of HR 3099 is shown in Fig. 5, together
with the identification of the stronger lines; in this Fig. 
the synthetic spectrum computed for 
(T$_{\rm eff}$, log g, [Fe/H], v$_{\rm t}$) = 
(3750 K, 1.5, 0.0, 2.0 km.s$^{-1}$)
is also superimposed.
Notice that this temperature is higher than the one indicated in Table 1
(derived from spectral type) for HR 3099, suggesting that this star 
could be classified as M2 rather than M6.

On the basis of our higher resolution spectra and our spectrum
synthesis computations, we confirm the presence of contaminating lines
due to atomic lines of FeI, VI and ZrI and a TiO bandhead (Fig. 5).
For giants of T$_{\rm eff}$ $<$ 3600 K the TiO band dominates the blend. 

\bigskip 

\noindent {\bf 5. Discussion}
\bigskip
\noindent The response of the NaI doublet to the variation of surface gravity 
is pronounced (Fig. 4b). We conclude that the doublet can be used
as a dwarf-giant discriminator.
As shown in Figs. 4a and 4b, both definitions of the NaI index
(EW$_1$ and EW$_2$) can discriminate between dwarfs and giants.
We note however that EW$_1$ includes the bottom of the TiO (0,2)
bandhead; in a giant, the total EW$_1$ will be more due to TiO
than to the NaI feature, as shown in Fig. 3.
The fact of measuring TiO rather than the NaI lines makes
EW$_1$ slightly sensitive to temperature (and metallicity) for giants, and it
introduces a difficulty to disentangle gravity from 
temperature effects.
\par For these reasons we propose the index EW$_2$
 which adopts an
integration of the line profile in the range $\lambda\lambda$ 8172-8197
${\rm \AA}$; we also suggest the use of
a long wavelength continuum window different
from the one previously used (we suggest $\sim\lambda$ 8234 ${\rm \AA)}$,
but in fact both continua definitions give a similar dwarf/giant
discrimination.

Several authors suggested that the enhancement of the NaI feature in the
integrated spectra of some galaxies is due to a higher metallicity, rather
than due to a dwarf dominated population. 
The NaI index, as defined in the literature, is somewhat affected by TiO
bands:
high metallicity stars are cooler, favoring the strengthening of
TiO bands, so that in an indirect way, an enhancement of the NaI feature
could be assigned to metallicity.

\bigskip
\noindent {\bf Acknowledgements.}
\bigskip
\noindent The authors are indebted to B. Plez and F. Allard for kindly
making available their model atmospheres.
The calculations were carried out in DEC Alpha 3000/700
workstation provided by Fapesp. 
RPS and AM acknowledge Fapesp PhD fellowships n$^o$
93/2177-0 and 91/2100-1. 
Partial financial support from CNPq is also acknowledged.

\vfill\eject
\noindent {\bf References}
\bigskip
\ref Allard, F., Hauschildt, P.H.: 1995, ApJ, 445, 433 
\ref Alloin, D., Bica, E.: 1989, A\&A, 217, 57 (AB89)
\ref Barbuy, B.: 1982, PhD thesis, Universit\'e de Paris VII
\ref Bessell, M.S.: 1991, AJ, 101, 662
\ref Boroson, T.A., Thompson, I.B.: 1991, AJ, 101,111
\ref Cohen, J.G.: 1978, ApJ, 221, 788
\ref Couture, J., Hardy, E.:1993, ApJ, 406, 142
\ref Davis, S.P., Phillips, J.G.: 1963, {\it The red system
(A$^2$$\Pi$ - X$^2$$\Sigma$) of the CN molecule}, Univ. California Press
\ref Delisle, S., Hardy, E.: 1992, AJ, 103, 711 
\ref Erdelyi-Mendes, M., Barbuy, B.: 1991, A\&A, 241, 176
\ref Faber, S.M., French, H.: 1980, ApJ, 235, 405 (FF80)
\ref Fluks, M.A., Plez, B., Th\'e, P.S., De Winter, D., 
Westerlund, B.E., Steeman, H.C.: 1994, A\&AS, 105, 311
\ref Kurucz, R.: 1992, in {\it The Stellar Populations of Galaxies},
IAU Symp. 149, eds. B. Barbuy \& A. Renzini, Kluwer Acad. Press, 225
\ref Milone, A., Barbuy, B.: 1994, A\&AS, 108, 449
\ref Moore, C.E., Minnaert, M.G., Houtgast, J.: 1966, NBS
Monograph n$^o$ 61, Washington
\ref Persson, S.E., Cohen, J.G., Sellgren, K., Mould, J., Frogel, J.A.:
1980, ApJ, 240, 779
\ref Plez, B., Brett, J.M., Nordlund, A.: 1992, A\&A, 256, 551
\ref Spinrad, H., Taylor, B.J.: 1971, ApJS, 22, 445 (ST)
\ref Xu, Z., V\'eron-Cetty, M.-P., V\'eron, P.: 1989, A\&A, 211, L12
\ref Whitford, A.E.: 1977, ApJ, 211, 527
\ref Wiese, W.L., Smith, M.S., Miles, B.M.: 1969, {\it Atomic
Transition Probabilities. II. Sodium Through Calcium},
NBS n$^o$ 22

\vfill\eject
\baselineskip=15pt

\noindent {\bf Table 1 - Log-book of observations}

$$\vbox{\halign to \hsize{\tabskip 1em plus2em$
#\hfil$&
$ #\hfil$&
$ #\hfil$&
 $#\hfil$&
$ #\hfil$&
 $#\hfil$&
 $ #\hfil\quad$&
$ #\hfil $ &
 $ #\hfil$&
$# \hfil $$\cr
\noalign{\hrule\vskip 0.2cm}
{\rm star} &
\hidewidth {\rm V} \hidewidth  &
\hidewidth {\rm Sp.T.} \hidewidth &
\hidewidth {\rm T_{eff}} \hidewidth &
\hidewidth {\rm star} \hidewidth &
\hidewidth {\rm V} \hidewidth &
\hidewidth {\rm Sp.T.} \hidewidth & 
\hidewidth {\rm T_{eff}} \hidewidth & \cr
\noalign{\vskip 0.2cm}
  \hidewidth {\rm Giants} \hidewidth  & &
\hidewidth  \hidewidth & &
\hidewidth {\rm Dwarfs} \hidewidth
\hidewidth  \hidewidth & \cr
\noalign{\vskip 0.2cm}
\noalign{\hrule\vskip 0.2cm}
\noalign{\vskip 0.2cm}
{\rm HR\;1693} & 5.68 & {\rm M6} &3300& {\rm GL\;190}&10.3 &{\rm M4} &3250 & \cr
{\rm HR\;2156} &6.95 & {\rm M6} &3300 & {\rm GL\;229} &8.14 & {\rm M1e} &3630 & \cr
{\rm HR\;2168} &5.31 & {\rm M2}&3810 & {\rm GL\;273} &9.85 & {\rm M3.5}&3220 & \cr
{\rm HR\;2469} &5.19 & {\rm M0}&3900 & {\rm GL\;285} &11.2 &{\rm M4.5e}&3140 & \cr
{\rm HR\;2802} &5.87&  {\rm M4}&3570 & {\rm GL\;357} &10.92 & {\rm M3}&-- & \cr
{\rm HR\;3099} &6.33 & {\rm M6}&3300 & {\rm GL\;406} &13.45 & {\rm M6}&-- & \cr
{\rm HR\;3816} &6.1 & {\rm M6-7epv}&3220 & {\rm GL\;493.1} &13.4 & {\rm M5}&3090 &  \cr
{\rm HR\;3793} &5.88 & {\rm M2} &3810 & {\rm GJ\;1142A} &12.56 &{\rm M6}&-- & \cr
{\rm HR\;4045} &6.3 & {\rm M4-5}&3500 & & & & & \cr
{\rm HR\;4267} &5.81 & {\rm M5.5} &3370 & & & & & \cr
\noalign{\vskip 0.01cm}  \cr}
\hrule}$$

\noindent {\bf Table 2 - Stellar parameters adopted}

$$\vbox{\halign to \hsize{\tabskip 1em plus2em$
#\hfil$&
$ #\hfil$&
$ #\hfil$&
 $#\hfil$&
 $ #\hfil\quad$&
$ #\hfil $ &
 $ #\hfil$&
$# \hfil $$\cr
\noalign{\hrule\vskip 0.2cm}
{\rm models} &
\hidewidth {\rm T_{eff}} \hidewidth  &
\hidewidth {\rm log\;g} \hidewidth &
\hidewidth {\rm [Fe/H]} \hidewidth & \cr
\noalign{\vskip 0.1cm}
\noalign{\hrule\vskip 0.2cm}
\noalign{\vskip 0.2cm}
{\rm Kurucz} & 3500\;{\rm to}\;5000 & 1.0,4.5 & +0.5\;{\rm to}\;-2.0 & \cr
{\rm Plez} & 2750\;{\rm to}\;3400 & -0.5\;{\rm to}\;1.5 & 0.0 & \cr
{\rm Allard} & 3500\;{\rm to}\;2700 & 5.0 & +0.5\;{\rm to}\;-3.0 & \cr
\noalign{\vskip 0.01cm}  \cr}
\hrule}$$

\noindent {\bf Table 3 - Wavelengths defining continua and NaI index}

$$\vbox{\halign to \hsize{\tabskip 1em plus2em$
#\hfil$&
$ #\hfil$&
$ #\hfil$&
 $#\hfil$&
 $ #\hfil\quad$&
$ #\hfil $ &
 $ #\hfil$&
$# \hfil $$\cr
\noalign{\hrule\vskip 0.2cm}
{\rm } &
\hidewidth {\rm blue\; cont.} \hidewidth  &
\hidewidth {\rm NaI\; index} \hidewidth &
\hidewidth {\rm red\;cont.} \hidewidth & 
\hidewidth {\rm designation} \hidewidth & \cr
\noalign{\vskip 0.2cm}
\noalign{\hrule\vskip 0.2cm}
\noalign{\vskip 0.2cm}
{\rm FF80/AB89} & 8169-8171 & 8172-8209 & 8209-8211 & {\rm EW_1} & \cr
{\rm present\;work} & 8171.5-8172. & 8172-8197 & 8233.5-8234.2 & {\rm EW_2} & \cr
\noalign{\vskip 0.01cm}  \cr}
\hrule}$$
\baselineskip=21pt

\vfill\eject
\vtop to 1.0 true cm {}
\centerline {\bf FIGURE CAPTIONS}
\bigskip
\noindent {\bf Figure 1} - Observed spectra 
in order of decreasing temperatures from top to bottom for (a) giants:
HR2469 (M0), HR4045 (M4-5), HR3099 (M6 - see Sect. 4.1),
HR4267 (M5.5), HR2156 (M6), HR1693 (M6), HR3816 (M6-7);
(b) dwarfs: GL229 (M1), GL357 (M3), GL190 (M4),
 GL493.1 (M5), GL406 (M6).
\bigskip
\noindent {\bf Figure 2} - GL 493.1: observed spectrum (dashed line)
and synthetic spectrum (solid line)
computed with (T$_{\rm eff}$, log g, [Fe/H], v$_{\rm t}$) = 
(3100 K, 4.5, 0.0, 1.0 km.s$^{-1}$).

\bigskip
\noindent {\bf Figure 3 } - Synthetic spectra for
a dwarf and a giant, where the continuum defined in the literature
(lower line) and our adopted continuum  (upper line) are indicated.

\bigskip
\noindent {\bf Figure 4} - 
 Equivalent widths of the NaI feature, for giants and dwarfs, 
 adopting the NaI index definitions
(a) EW$_1$ (present work) and (b) EW$_2$ (literature).

\bigskip
\noindent {\bf Figure 5} - 
High resolution spectrum of HR3099 (dotted line), 
synthetic spectrum computed for 
(T$_{\rm eff}$, log g, [Fe/H], v$_{\rm t}$) = (3750, 1.5, 0.0, 2.0)
(solid line). The main lines are identified.

\vfill
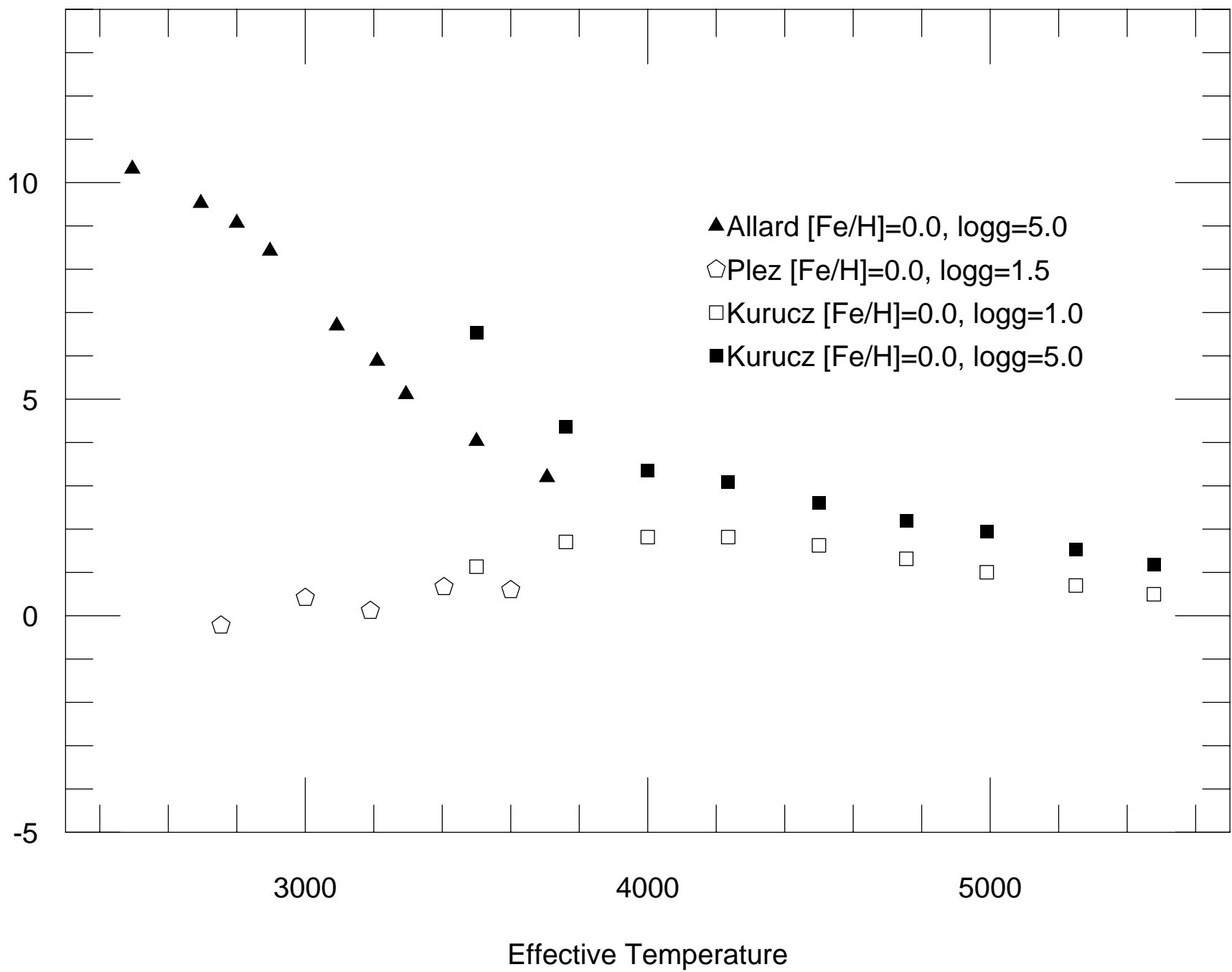
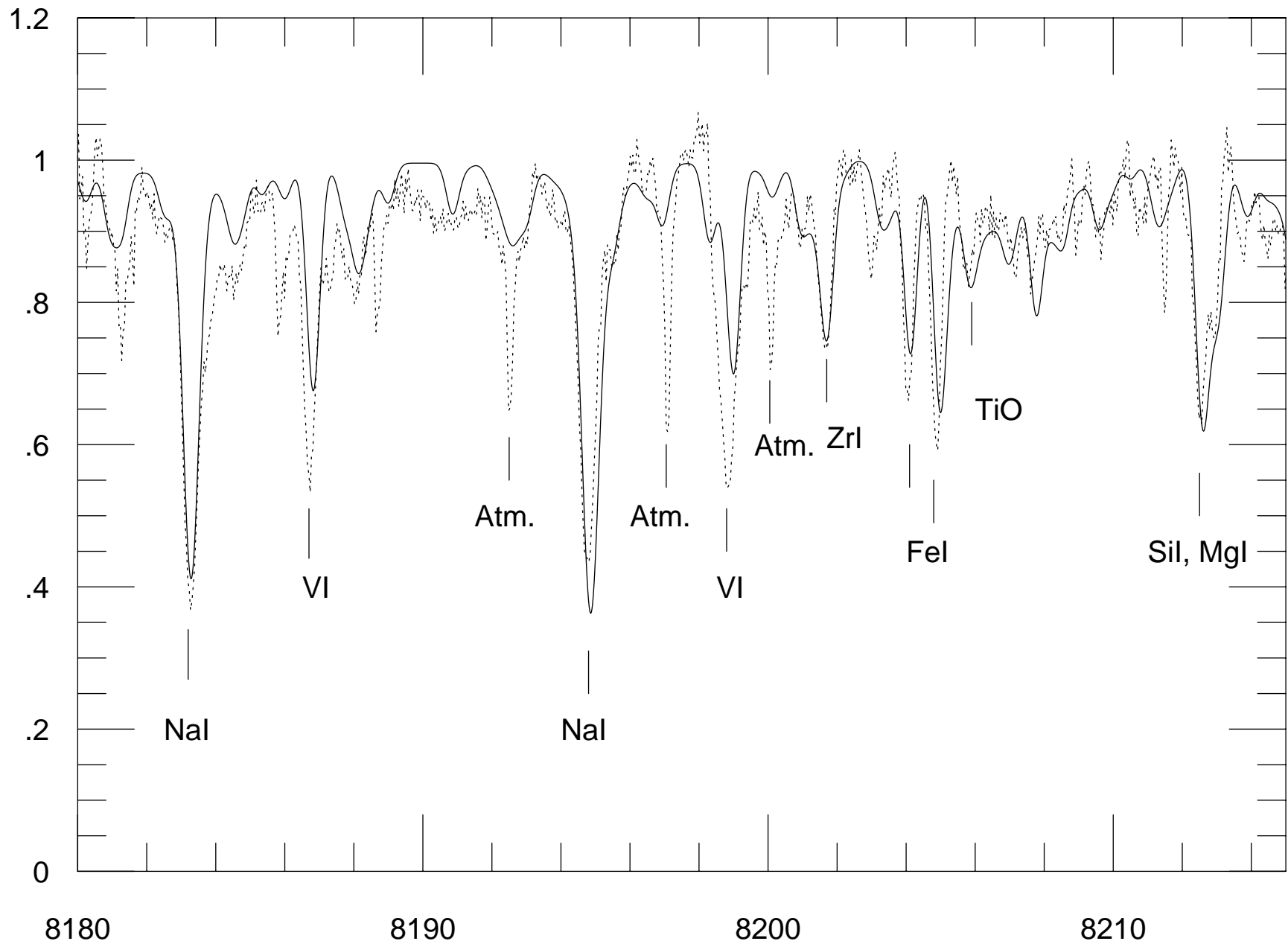
\bye